\documentclass[sn-mathphys,Numbered]{sn-jnl}

\usepackage{graphicx}%
\usepackage{multirow}%
\usepackage{amsmath,amssymb,amsfonts}%
\usepackage{amsthm}%
\usepackage{mathrsfs}%
\usepackage[title]{appendix}%
\usepackage{xcolor}%
\usepackage{textcomp}%
\usepackage{manyfoot}%
\usepackage{booktabs}%
\usepackage{algorithm}%
\usepackage{algorithmicx}%
\usepackage{algpseudocode}%
\usepackage{listings}%

\theoremstyle{thmstyleone}%
\theoremstyle{thmstyletwo}%
\theoremstyle{thmstylethree}%

\begin{document}

\title[MACHO constraints on PBH from Gaia]{Reanalysis of the MACHO constraints on PBH in the light of Gaia DR3 data}

\author*[1]{\fnm{Juan} \sur{Garc\'ia-Bellido}}\email{juan.garciabellido@uam.es}

\author[2]{\fnm{Michael} \sur{Hawkins}}\email{mrsh@roe.ac.uk}
\equalcont{These authors contributed equally to this work.}

\affil*[1]{\orgdiv{Instituto de F\'isica Te\'orica UAM/CSIC}, \orgname{Universidad Aut\'onoma de Madrid}, \orgaddress{\street{Nicol\'as Cabrera 13}, \city{Madrid} \postcode{28049}, \country{Spain}}}

\affil[2]{\orgdiv{Institute for Astronomy}, \orgname{University of Edinburgh, Royal Observatory}, \orgaddress{\street{Blackford Hill}, \city{Edinburgh}, \postcode{EH9 3HJ}, \country{UK}}}

\abstract{The recent astrometric data of hundreds of millions of stars from Gaia DR3 has allowed a precise determination of the Milky Way rotation curve up to 28~kpc. The data suggests a rapid decline in the density of dark matter beyond 19~kpc. We fit the whole rotation curve with four components (gas, disk, bulge and halo) and compute the microlensing optical depth to the Large Magellanic Cloud. With this model of the galaxy we reanalyse the microlensing events of the MACHO and EROS-2 Collaborations.  Using the published MACHO efficiency function for the duration of their survey, together with the rate of expected events according to the new density profile, we find that the Dark Matter halo could be composed up to 100\% of massive compact halo objects for any mass between $10^{-4}$ to $100~M_\odot$, except a broad range around $0.01~M_\odot$, where it cannot be larger than $\sim20~\%$.  For the EROS-2 survey, using a modified efficiency curve for consistency with the MACHO analysis, we also find compatibility with a 100\% MACHO halo, but with a tighter constraint around $0.001~M_\odot$ where the halo fraction cannot be larger than $\sim12~\%$. This result assumes that MACHOs all have the same mass. If these were distributed in an extended mass function like that of the Thermal History Model, the constraints are weakened, allowing 100\% of all DM in the form of Primordial Black Holes.}


\maketitle

\section{Introduction}\label{introduction}

The idea that a large number of gravitationally collapsed objects might have formed in the early Universe as a result of density fluctuations was first proposed by \citep{Hawking:1971ei}.  Such objects soon became known as primordial black holes (PBH), with the idea generally attributed to \citep{Chapline:1975ojl} that they might account for the dark matter (DM), see also~\cite{Carr:2023tpt,LISACosmologyWorkingGroup:2023njw} for recent reviews. A simple model for large density fluctuations generated during inflation that could give rise to PBH comprising all of the DM was soon proposed by~\citep{Garcia-Bellido:1996mdl}. A further important development was the suggestion by~\citep{Jedamzik:1996mr} that the QCD phase transition should lead to a peak in black hole production of around a solar mass.  The detection of black holes of this mass present formidable difficulties, as any radiation from accretion would be negligible, and gravitational interaction with luminous bodies would be hard to identify unambiguously.  However, a suggestion by~\citep{Paczynski:1985jf} that solar mass compact bodies making up the dark matter might be detected from the occasional microlensing of stars in the Large Magellanic Cloud (LMC) was being followed up by a large scale photometric monitoring of several million stars in the Magellanic Clouds~\citep{MACHO:1995udp}.  The results of this survey are well known~\citep{MACHO:2000qbb}, and include the detection of some 13 unambiguous microlensing events, identified as solar mass compact bodies.  These detections were far in excess of those expected from any known stellar population.

In order to estimate the contribution of these bodies to the mass budget of the halo, two further steps are necessary.  Firstly, the detection efficiency must be estimated.  This relates the number of observed microlensing events to the total number which could have been detected according to the search criteria.  There is much uncertainty about this figure \citep{Hawkins:2015uja}, highlighted by differences and inconsistencies between the results of other surveys such as EROS \citep{EROS-2:2006ryy} and OGLE \citep{Wyrzykowski:2011tr}.

The third major requirement for estimating the halo fraction in compact bodies is a reliable model for the Milky Way, incorporating mass estimates for the major components of halo, bulge, disc and gas.  Of particular significance for determining these parameters is the rotation curve of the Milky Way, which is the main subject of this paper.  Early attempts to measure the Galactic rotation curve where confined to measures of the CO molecular line and HI using the tangent point detection method \citep{Clemens:1985}.  This approach gave unreliable estimates of the rotation speed even within the orbit of the sun, and provided virtually no useful information at larger Galactocentric radii.  Faced with this difficulty, the MACHO collaboration opted for a different approach, noting that the rotation curves of nearby galaxies which were relatively easy to measure were for the most part flat.  On this basis they considered a number of galaxy templates, but in their analysis favoured a heavy halo model with a flat rotation curve.  This assumption lead to the conclusion that although the population of compact bodies which had been detected from their microlensing signal could not be accounted for from known stellar populations, they were insufficient to make up the dark matter component of the Galactic halo.

In the years following the publication of the MACHO results \citep{MACHO:2000qbb}, a number of groups set out to make direct measurements of the the Milky Way rotation curve using observations of velocity dispersion in samples of halo star populations.  With the publication of the first of these results \citep{Xue:2008nmx,Sofue:2013kja,Bhattacharjee:2013exa}, it became clear that the flat rotation curve model adopted by the MACHO collaboration \citep{MACHO:1995udp,MACHO:2000qbb} was not consistent with the new data \citep{Hawkins:2015uja}.  This result was soon confirmed by \citep{Calcino:2018mwh} from new measurements of the rotation curve by \citep{Huang:2016}.

Although the new Milky Way rotation curves were clearly incompatible with the standard flat rotation model adopted by \citep{MACHO:2000qbb}, they were nonetheless noisy, with unexpected departures from a smooth profile which suggested possible artefacts in the data.  The situation has recently been transformed with the publication of a new high signal-to-noise Milky Way rotation curve based on data from the {\it Gaia} DR3 data release \citep{Ou:2023adg}, and the detection of a rotational velocity decline \citep{Jiao:2023aci} implying an effective cut-off in the mass distribution of the Galaxy~\cite{Wang:2023,SylosLabini:2023}. This has significant implications for any limits to a hypothetical population of PBH constituting the dark matter halo.

\begin{figure}[h]%
\centering\hspace*{-2mm}
\includegraphics[width=\textwidth]{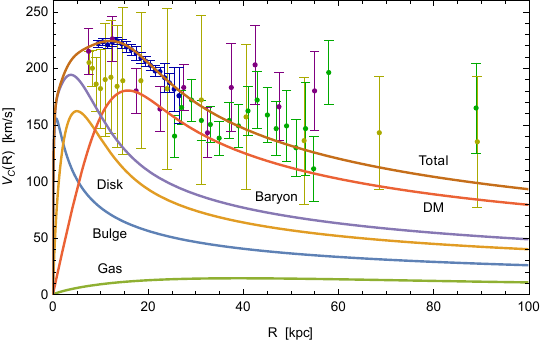}
\caption{
The rotation velocity curve of the Milky Way as a funtion of distance from the center in kpc. We plot the {\it Gaia} DR3 data (blue) from Ref.~\citep{Jiao:2023aci}, together with data from Ref.~\cite{Bhattacharjee:2013exa,Sofue:2013kja,Xue:2008nmx} (green and purple), and the best fits to the four galactic components (gas, bulge, disk and halo). 
}
\label{fig:MWRV}
\end{figure}

\section{Results}\label{results}

The constraints on the possible MACHO abundance in the Milky Way dark matter halo depends extraordinarily on accurate information and modelling of our galaxy's rotation curve, which is used to determine the density profile and thus the probability of microlensing to distant stars. The rotation curve outside the solar radius ($R_\odot \simeq 8.5$~kpc) requires accurate knowledge of the 3D motion of the Sun and the distance to all available sources. Previous generation surveys used a large sample of stars and determined their distance using the distance modulus. For example, the compilation by~\citep{Huang:2016} uses around 22,000 stars out to a distance of 100 kpc. The recent revolution on galactic determination of distances has come via the astrometric survey {\em Gaia} and its third data release (DR3)~\citep{DR3}, which determined the 6D phase-space (parallaxes and proper motions) of over 30,000 RGB stars out to 30 kpc, which has allowed astronomers to determine the circular rotation velocity of the galaxy~\citep{Ou:2023adg,Jiao:2023aci}, assuming spherical symmetry.

\subsection{Milky Way Rotation Curve}

We parametrize the Milky Way rotation curve as a four component model of the galaxy with distinct bulge, disk, gas and halo components. Each component is characterized by a mass profile $M_i(R)$ and a corresponding circular rotation velocity $v_i^2(R)=GM_i(R)/R$. The total mass at a given radial distance from the Galactic Center is the sum of all the components $M_{\rm tot}(R)=M_b(R) + M_d(R) + M_g(R) + M_h(R)$, and thus the circular rotation curve velocity at galactocentric radius $R$ is given by $v_{\rm tot}^2(R)=GM_{\rm tot}(R)/R = v_b^2(R) + v_d^2(R) + v_g^2(R) + v_h^2(R)$. We describe in the following subsections the different component models. For all components, we quote the best fit values of the model parameters, as determined by~\citep{Jiao:2023aci}.

\subsubsection{Bulge}

We parametrize the bulge component with a Hernquist profile
\begin{equation}
\Phi(r) = - \frac{GM_b}{r+r_b} 
\ \Longrightarrow \ 
M(r) = \frac{M_b r^2}{(r+r_b)^2} 
\ \Longrightarrow \ 
v(r) = \frac{\sqrt{GM_b\,r}}{(r+r_b)}
\label{eq:phi-bulge}
\end{equation}
with $M_b = 1.55\times10^{10}\,M_\odot$ and $r_b = 0.70$ kpc. 

\subsubsection{Disk}

The disk is parametrized with an exponential profile, with a radial distance $r$ along the disk and a width $z$ orthogonal to the plane,
\begin{eqnarray}
    \rho(R) &=& \rho_0 \exp\left[-\frac{r}{L} - \frac{|z|}{H}\right]  
    \ \Longrightarrow \ 
v(R) = \sqrt{\frac{GM_d}{L}}\,D(x) \\[2mm]
    D(x) &=& \frac{x}{\sqrt2}\left[
    I_0\left(\frac{x}{2}\right)
    K_0\left(\frac{x}{2}\right) - 
    I_1\left(\frac{x}{2}\right)
    K_1\left(\frac{x}{2}\right)
    \right]^{1/2}
\label{eq:rho-disk}
\end{eqnarray}
with $x=R/L$, $M_d = 3.65\times10^{10}\,M_\odot$, $L = 2.35$ kpc and $H = 0.14$ kpc. The $I_n$ and $K_n$ are modified Bessel functions of the first and second kind respectively.

\subsubsection{Gas}

The gas component is also characterized with an exponential profile
\begin{equation}
    \rho(R) = \rho_0 \exp\left[-\frac{r}{R_g} - \frac{|z|}{z_d}\right]  
    \ \Longrightarrow \ 
v(R) = \sqrt{\frac{GM_g}{R_g}}\,D(x) 
\label{eq:rho-gas}
\end{equation}
with $x=R/R_g$, $M_g = 0.82\times10^{10}\,M_\odot$, $R_g = 18.14$ kpc and $z_d = 0.52$ kpc.

\subsubsection{Halo}

The halo component is best characterized by the Einasto profile of index $n$,
\begin{equation}
    \rho(R) = \rho_0 \exp\left[
    -\left(\frac{R}{R_h}\right)^{1/n} \right]  
    \ \Longrightarrow \ 
    M(R) = 4n\pi\rho_0\,R_h^3\times \Gamma\left[
    3n,\left(\frac{R}{R_h}\right)^{1/n}\right]
\label{eq:rho-halo}
\end{equation}
with $\rho_0 = 0.01992\,M_\odot/{\rm kpc}^3$, $R_h = 11.41$ kpc and $n = 0.43$. Here $\Gamma[n,x]$ stands for the incomplete Gamma function.

\subsubsection{Total}

We have shown in Fig.~\ref{fig:MWRV} the rotation velocity curve of the Milky Way as a function of radius. The four-component model (gas, bulge, disk and halo) is enough to describe the {\it Gaia} DR3 data~\citep{Jiao:2023aci}. We note that the new data (in blue) is significantly better determined than previous surveys (in green and purple), from Refs.~\cite{Bhattacharjee:2013exa,Sofue:2013kja,Xue:2008nmx}. We have computed the statistical significance of the whole data set (from Jiao, Bhattacharjee, Sofue and Xue) and found that the best fit model for the rotation curve derived from just the Jiao {\it et. al.} data does not differ within the error bars from the global best fit, and has a $\chi^2/{\rm d.o.f.}=1.23$, which is a near perfect fit. We will thus use this model to reanalyse the MACHO constraints of the Milky Way.


\begin{figure}[h]%
\centering\hspace*{-2mm}
\includegraphics[width=0.475\textwidth]{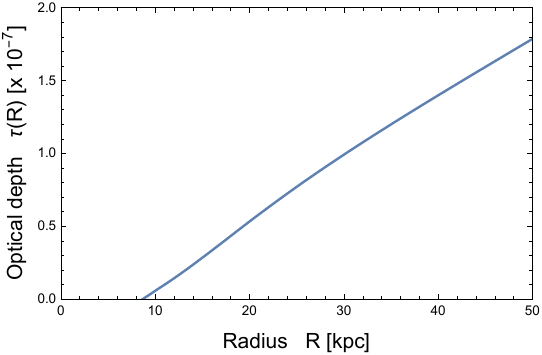}
\hspace{2mm}
\includegraphics[width=0.495\textwidth]{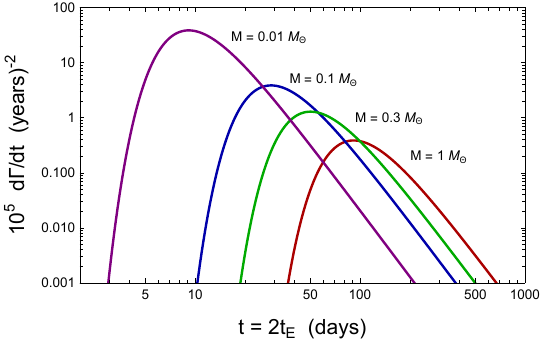}
\caption{{\em Left}: The microlensing optical depth to a source at a distance $R$ from our location near the sun, at 8.5 kpc from the center of our galaxy. {\em Right}: The differential rate of microlensing events as a function of crossing time $\hat t = 2t_{\rm E}$ (in days). We show three cases, for MACHOs of $1~M_\odot$ (red curve), $0.3~M_\odot$ (green curve), $0.1~M_\odot$ (blue curve) and $0.01~M_\odot$ (purple curve).}\label{fig:tau}
\end{figure}

\section{The microlensing optical depth and rate of events}\label{opticaldepth}

MACHOs can be detected by their gravitational influence as they pass close to the line of sight of a distant star~\citep{Paczynski:1985jf}. The source will appear distorted into multiple images. If the lens and source were perfectly aligned, the source appears as an Einstein ring of radius given by
\begin{equation}
    r_E^2(x) = \frac{4GML}{c^2}\cdot x(1-x)
\end{equation}
where $M$ is the mass of the lens, $L$ is the distance between the observer and the source, and $x=R/L$. Since perfect alignment is very unlikely, most of the times the lens will create multiple images of the source, which cannot be resolved and thus their light adds up, producing a characteristic brightening of the source that follows (as a function of time) what is called a Paczynski curve.

The microlensing optical depth $\tau$ is defined as the number of compact lenses within a tube of Einstein radius and length $L$. Is can thus be computed from the density along the line of sight
\begin{equation}
    \tau = \frac{4\pi G}{c^2}\int_{x_\odot}^1 \,\rho(Lx)\,x(1-x)\,dx =
    \frac{G}{c^2L}\int_{x_\odot}^1 \,M(x)\,(x-\ln x)\,dx
    \label{eq:tau}
\end{equation}
where we have used $M'(x) = 4\pi L\,x^2\rho(Lx)$ with $R=Lx$ and $x_\odot = R_\odot/L$, and we have integrated by parts. We plotted the microlensing optical depth in Fig.~\ref{fig:tau}a. We see that it reaches values of order $\tau\sim1.8\times10^{-7}$ at a distance of 50 kpc to the LMC.

The rate of microlensing events, $\Gamma=4/\pi\cdot\tau/\langle\hat t\rangle$, is essentially the optical depth $\tau$ over the average crossing time, $\hat t = 2t_E$, that a star in the background  (e.g. the LMC) takes to cross the lens' Einstein ring. The differential rate per unit time again depends on the distance to the source and the density profile,
\begin{equation}
    \frac{d\Gamma}{d\hat{t}}(M) = \frac{32L}{\hat{t}^4 v_c^2 M} \int_{x_\odot}^1 \,\rho(x)\,r_E^4(x)\,\exp\left[-\frac{4r_E^2(x)}{\hat{t}^2 v_c^2}\right]\,dx 
    \label{eq:dG}
\end{equation}
We show in Fig.~\ref{fig:tau}b the differential rate of microlensing events as a function of crossing time $\hat t$ for three cases, MACHOs of 1, 0.3 and 0.01 solar masses.

\begin{figure}[h]%
\centering\hspace*{-2mm}
\includegraphics[width=0.485\textwidth]{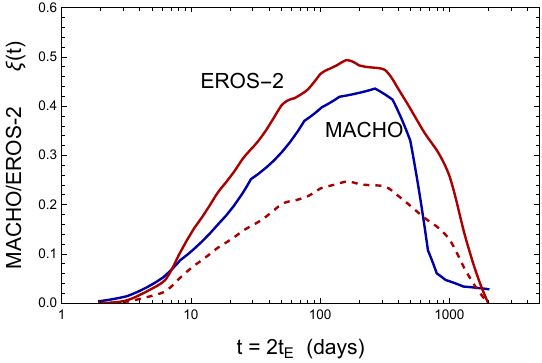}
\hspace{2mm}
\includegraphics[width=0.485\textwidth]{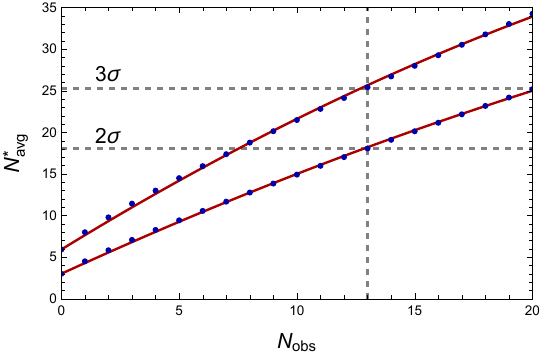}
\caption{{\em Left}: The efficiency curves for MACHO (blue curve), EROS-2 (red curve) and 50\% less efficient EROS-2 (red dashed curve), as a function of crossing time in days. {\em Right}: The bound $N_{\rm obs}^*$ on the average number of expected events as a function of the observed events $N_{\rm obs}$, for a Poisson distribution, at $2\sigma$ (95\%) and $3\sigma$ (99.73\%) confidence. The dashed lines correspond to $N_{\rm obs}=13$.}
\label{fig:eff}
\end{figure}

In order to compute the number of expected microlensing events, we need to integrate over the experiment's efficiency function $\xi(\hat t)$,
\begin{equation}
    N_{\rm exp}(M) = E\,\int_0^\infty 
    \frac{d\Gamma}{d\hat{t}}(M)\,\xi(\hat t)\,d\hat t
    \label{eq:Nexp}
\end{equation}
where $E$ is the total exposure time in units of the number of monitored stars times the duration of the survey in years. In the case of the MACHO survey, which lasted for 5.7 years, they estimated $E=6.12\times10^7$ objects $\cdot$ years~\citep{Bennett:2005at}. In the case of the EROS-2 survey, they estimated $E=3.8\times10^7$ objects $\cdot$ years~\citep{EROS-2:2006ryy}. The efficiency function describes the ability of a given survey to identify microlensing events over a range of times $\hat t$. If a survey only runs for a short period of time, it will not be able to detect long duration events. On the other hand, if the cadence of the survey is too sparse, it will not be able to detect short duration events. Since the duration of the event depends on the mass of the lens, its distance to the observer and source, and the transverse velocity of the lens, these limitations of the survey reflect on the range of masses that it can potentially detect. The MACHO and EROS-2 efficiencies are still a matter of debate~\citep{Hawkins:2015uja}, and we will postpone their discussion to the next sections. We will take for the moment the published ones, see Fig.~\ref{fig:eff}a.

\subsection{Microlensing detection Efficiency}

In microlensing surveys the number of events detected as satisfying certain criteria or `cuts' as a fraction of the actual number of events which satisfied those cuts is conventionally known as the detection efficiency $\xi(\hat{t})$, where $\hat{t}$ is the timescale of the microlensing event.  It is worth noting that in a crowded field the detection efficiency is also a strong function of the brightness of the source.  The brightness is not related to $\hat{t}$, which is a function of the mass of the lens.  In addition, it is clear that the detection efficiency depends crucially on the choice and severity of the cuts for a particular sample of candidate microlensing events.  These distinctions are important, as one would not for example expect the detection efficiency for a bright star sample to be the same as for an `all star' sample.  The focus on $\hat{t}$ would appear to originate from interest in improvements in detection efficiency as the MACHO collaboration extended the length of their monitoring programme.

In a sparse field, the calculation of the detection efficiency is fairly straightforward, and can be done analytically, based on the surface density of the sources (stars) and the adopted impact parameter.  However, the dense star fields of the Magellanic Clouds present a completely different order of problem.  To give an idea of the difficulties involved, {\it Hubble Space Telescope} (HST) follow-up of the MACHO detections provides a useful insight.  In Figure 3 of Ref.~\citep{Alcock:2000zv} the MACHO point spread function is superimposed upon the HST image of the field around event LMC-4.  It can be seen that the MACHO point spread function contains no less than 5 stars resolved in the HST image.  It is clear that any event occurring in one of those stars will have to be very bright to be recorded as a detection, and the light curve will be distorted in colour and shape by the presence of the non-varying stars.  These effects are discussed in detail in Ref.~\cite{Hawkins:2015uja}.

To address the difficulties associated with estimating the detection efficiency in the crowded star fields, the MACHO collaboration opted for a variation on the Monte Carlo method which, rather than assigning random values to input parameters, superimposed simulated microlensing events onto a selection of their light curves, and then applied their cuts to assess the probability of detection.  In an attempt to avoid the difficulties associated with crowded fields, the EROS collaboration opted to restrict their EROS-2 survey to bright stars with a magnitude cut-off at around $R < 18.5$, see Ref.~\citep{EROS-2:2006ryy}.  Given that when the MACHO survey was complete, only 2 of their 17 microlensing events were bright enough to be included in the EROS-2 bright sample, this was perhaps an unfortunate choice.  However, EROS observations commenced before the completion of the MACHO observations in December 1999, giving an overlap of three and a half years in part of the area of sky covered by the MACHO survey, and 3 of the 17 MACHO events occurred in this overlap envelope.  One of these was below the EROS detection limit, and is not recorded in their measurements.  The other two were detected but not accepted as candidates since, apart from being too faint for the bright star sample, they failed to pass the EROS-2 cuts for the full sample \citep{EROS-2:2006ryy}.

Despite the advantage of greater brightness, there are reasons why luminous stars, which are most likely to be red giants, may be less likely to be observed as microlensing candidates.  For a $0.7~M_\odot$ lens (the most likely mass for a primordial black hole \citep{Byrnes:2018clq}), the Einstein radius for a Magellanic Cloud source star is $10^{9}$ km, which is sufficiently close to the radius of a giant star to significantly diminish the peak amplification in a microlensing event, as discussed in Ref.~\citep{Refsdal:1991}.  In addition, in a crowded star field the brightest sources are very likely to be the combination of several separate stars \citep{MACHO:2000bzs}, and the effect of any individual star being microlensed will be diluted by the non-varying light from its near neighbours.

Perhaps the final arbiter of detection efficiency should be the comparison of detections in overlapping fields by different surveys.  Returning to the overlap region between the MACHO and EROS-2 surveys mentioned above, with two expected events above the threshold of brightness (those observed by MACHO), a simple Poisson statistic suggests that the probability of zero detections by EROS-2 was around $P(0)= e^{-2} \simeq 0.14$, see Eq.~(\ref{eq:Poisson}). This assumes that the detection efficiencies were the same for both samples, as illustrated in Refs.~\citep{MACHO:2000qbb} and \citep{EROS-2:2006ryy}.  The difference in detection rate appears to be due to the comparative severity of the EROS-2 cuts which eliminated both MACHO detections.  This difference does not appear to have been accounted for in the detection efficiency calculation, which heavily depends on the nature of the cuts.  This can be put to the test by reducing the claimed EROS-2 detection efficiency.  For example, reducing the detection efficiency curve in Ref.~\citep{EROS-2:2006ryy} by a factor of two results in a Poisson probability of $P(0) = e^{-1} \simeq 0.37$, for zero detections in the overlap field.  Alternatively, one could maintain the same efficiency over the whole mass range and admit that instead of zero detections there were two observed microlensing events in EROS-2 survey. In that case the constraints are again weakened by a factor 2 since $N^*_{\rm avg}(N_{\rm obs} = 2)/N^*_{\rm avg}(N_{\rm obs} = 0)=5.83/2.99=1.95$, see Eq.~(\ref{eq:Nstar}).

This reduction essentially provides a solution to the longstanding puzzle of the  differences between MACHO and EROS-2 microlensing results, effectively ascribing it to differences in the cuts used to select microlensing candidates.  This in turn indicates a shortcoming in the ER0S-2 calculation of detection efficiency for reasons yet to be determined.  This can be remedied by reducing the EROS-2 detection efficiency curve illustrated in Figure 11 of Ref.~\citep{EROS-2:2006ryy} by a factor of two, which will be used for calculations in the remainder of this paper.  Although this is clearly not an exact figure, it should be closer to the true value than that published by the EROS-2 collaboration.

\section{Methods}\label{methods}

Once we have the number of expected microlensing events as a function of the mass of the lens (i.e. MACHOs), we can do some statistical analysis and infer the constraints on the halo mass fraction in MACHOs along the line of sight to the sources, e.g. stars in the LMC,
\begin{equation}
    C(M) = \frac{N_{\rm avg}}{N_{\rm exp}(M)} \leq \frac{N_{\rm avg}^*}{N_{\rm exp}(M)} 
    \label{eq:CM}
\end{equation}
where we expect that the number of {\em observed} events is Poisson distributed, given the {\em average} number of expected events,
\begin{equation}
    P(N_{\rm obs}) = 
    \frac{e^{-N_{\rm avg}}N_{\rm avg}^{N_{\rm obs}}}{N_{\rm obs}!} 
    \label{eq:Poisson}
\end{equation}
We want to find the value of $N_{\rm avg}$ such that we can be 95\% confident (corresponding to 2$\sigma$) that the true event rate is not larger than $N_{\rm avg}$. In other words, $P(N_{\rm obs})=1-0.95=0.05$ gives a dependence of $N_{\rm avg}\leq N_{\rm avg}^*$ on $N_{\rm obs}=z$, which is a transcendental equation
\begin{equation}
    N_{\rm avg}^* \simeq 3 + \ln\left(\frac{N_*^z}{z!}\right) \simeq
    3+1.3\,z - 0.01\,z^2
    \label{eq:Nstar}
\end{equation}
for which we have found a good polynomial fit for $N_{\rm obs}<20$, see Fig.~\ref{fig:eff}b. For example, in the EROS-2 survey, where no microlensing events were observed, we can be 95\% confident that the true even rate is not larger than 3 over the extent of the survey. If, on the other hand, we admit that there were two events observed in the EROS-2 survey, then we can be 95\% confident that the true event rate is not larger than 6. We take into account this factor of two increase in the upper limit of $C(M)$, see Eq.~(\ref{eq:CM}). We have also computed the corresponding values for 3$\sigma$ (or 99.73\% confidence), which gives a fit $N_{\rm avg}^*\simeq5.9 + 1.74 z - 0.017 z^2$, as shown in Fig.\ref{fig:eff}b. The 99.73\% c.l. constraints are significantly weaker, of course.

\begin{figure}[h]%
\centering
\includegraphics[width=\textwidth]{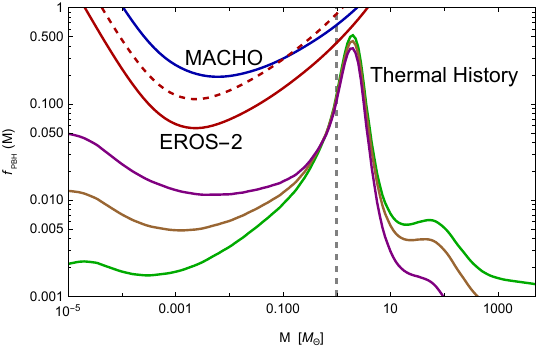}
\caption{The 95\% c.l. monochromatic constrains, as a function of the mass of the compact object, on the mass fraction of PBH in the galactic halo for the MACHO and EROS-2 collaborations assuming $N_{\rm obs}=13$ (blue curve), for the exposure time $E=6.12\times10^7$ star.yr, and for the EROS-2 coll., for  $E=3.8\times10^7$ star.yr and $N_{\rm obs}=0~(2)$ events (red (dashed) curves). We also overplot the extended mass function from the Thermal History Model~\citep{Carr:2019kxo}, for various spectral indices and running tilts. We clearly see that the halo can be composed of 100\% of dark matter in the form of PBH.}
\label{fig:fPBH}
\end{figure}

For a given value of $N_{\rm avg}^*$, the condition $C(M)\leq1$ constrains MACHOs of mass $M$ from contributing a 100\% of the Dark Mater halo. If $C(M)$ is found to be below one, then the mass fraction of the halo cannot be above $C(M)$ for such a value of the mass. These are the so-called monochromatic constraints, but can be extended to more complicated mass functions, like the ones associated with the Thermal History Model~\citep{Carr:2019kxo}, by simply integrating over the mass function $f(M)$,
\begin{equation}
    \int_0^\infty \frac{dM}{M} f(M) N_{\rm exp}(M) 
    \leq N_{\rm avg}^*
    \label{eq:fM}
\end{equation}
For mass functions normalized such that
$$\int_0^\infty \frac{dM}{M} f(M) = f_{\rm PBH},$$
we obtain the constraint
\begin{equation}
    f_{\rm PBH} \leq \left[\int_0^\infty \frac{dM}{M} \frac{f(M)}{C(M)} \right]^{-1}
    \label{eq:fPBH}
\end{equation}

\begin{figure}[h]%
\centering\hspace*{-2mm}
\includegraphics[width=\textwidth]{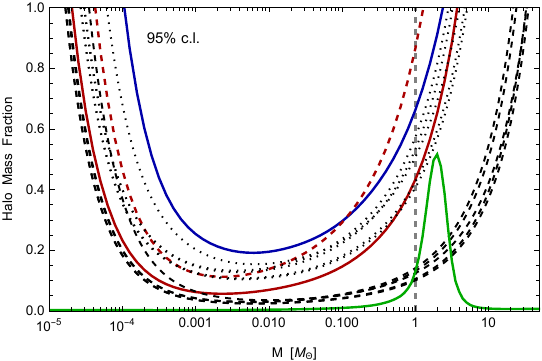}
\caption{
The comparison between the old constraints (MACHO=dotted, EROS-2=dashed) and the new ones from MACHO (blue) and EROS-2 (red) at 95\% c.l., with the dashed red curve corresponding to $N_{\rm obs}=2$ in EROS. We also plot the Thermal History Model (green). 
}
\label{fig:HMF}
\end{figure}

We have computed the monochromatic constraints on the mass fraction of the halo in the form of PBH (a specific type of MACHOs) for the new density profile of the Milky Way galaxy, for both MACHO and EROS-2 collaborations, as shown in Fig.~\ref{fig:fPBH}, for the MACHO coll. assuming $N_{\rm obs}=13$, for the exposure time $E=6.12\times10^7$~star.yr, and for the EROS-2 coll., for $N_{\rm obs}=0$ and $E=3.8\times10^7$~star.yr. In both cases, MACHO and EROS-2 results, the constraints allow $\sim 30$\% of all the halo for a monochromatic distribution with $M_{\rm PBH} \simeq 0.3~M_\odot$, while for the extended Thermal History Model mass function (red/green/purple lines in Fig.~\ref{fig:fPBH}), equation (\ref{eq:fPBH}) gives $f_{\rm PBH} \leq 1.9 ~(1.0)$ for MACHO (EROS-2) constraints, so we can easily accomodate a 100\% of the DM halo composed of PBH with the extended mass distribution of the Thermal History model.

\section{Discussion}\label{discussion}

The main purpose of this paper has been to use the Milky Way rotation curve from the latest {\it Gaia} data release to clear up some of the uncertainties surrounding the nature of the dark matter content of the Galactic halo.  The landmark microlensing survey of the Galactic halo by the MACHO collaboration detected some 13 events, most of which have now been unambiguously confirmed as the microlensing amplification of stars in the LMC by a population of compact bodies in the Galactic halo~\citep{Bennett:2005at}.  This population is far too large to be accounted for by any known stellar population, which has raised the question of whether they could make up the dark matter component of the halo.  There were however some problems with this hypothesis which we now discuss.

The most serious difficulty in accepting that the population of compact bodies constituted the dark matter was based on the belief that the number of observed microlensing events was too small to account for the expected number, based on models of the Galactic halo current at that time.  With hindsight, it seems strange that such a momentous discovery of a large population of unidentified compact bodies in the Galactic halo was not pursued with more vigour by the MACHO collaboration, whose main focus was whether the size of the population coincided with their modelling of the mass of the halo.  This may have been partly due to the distraction caused by the apparent failure of the EROS-2 collaboration, using new survey limits and cuts, to detect any microlensing events at all.

In this paper we have demonstrated that the new {\it Gaia} rotation curve tightly constrains any model of the Milky Way to one where there is no conflict between the microlensing event rate and a dark matter component of the Galactic halo made up of stellar mass compact bodies.  This is in contrast to the previous situation where the error bars on the published rotation curves where so large that it was hard to definitively rule out any Milky Way model.  Furthermore, the strong case for an extended mass function based on the thermal history of the universe removes any conflict between the MACHO microlensing detection rate and a dark matter halo entirely made up of primordial black holes.  We have also proposed a solution to account for the failure of EROS-2 to detect any microlensing events in their bright star sample.  This involves the conclusion that their more stringent cuts were not properly taken into account by their Monte Carlo efficiency estimates. Taking this into account gives consistent results that are compatible with those of MACHO collaboration.

\section{Conclusion}\label{conclusions}

The knowledge provided by the {\it Gaia} DR3 survey of hundreds of millions of stars in the Milky Way up to 30~kpc has revolutionized our way of measuring the rotation curve of our galaxy. This precise data has allowed us to construct a new four-component model of the galaxy which shows a rapid decline in density beyond 20~kpc, and at the same time consistent with previous measurements of the rotation curves thanks to their larger error bars. Since the most stringent constraints on the halo occupation fraction of compact objects comes from microlensing of stars towards the Large Magellanic Cloud, we have reanalysed those constraints in the light of the new model of the galaxy.

Our results suggest that both MACHO and EROS-2 collaboration observed number of events towards the LMC are compatible if we consider a factor two reduction in the efficiency of the EROS-2 survey or, alternatively, if we maintain the EROS-2 efficiency but include the two detected events in the overlapping fields common to both MACHO and EROS. In both cases, the constraints on Primordial Black Holes (as a particular realization of MACHOs) are significantly weakened, allowing for the totality of the Milky Way dark matter halo to be in the form of PBH of a given mass, except a broad range around $M_{\rm PBH} = 0.01~M_\odot$, where it cannot be larger than $\sim20~(12)\%$ for MACHO~(EROS) surveys. However, if we consider an extended mass function, like in the case of the Thermal History Model, then the constrains are further weakened and we can easily have 100\% of all the dark matter halo in the form of primordial black holes.

Our conclusions may gain further support when the {\it Gaia} collaboration releases their final astrometric data for stars out to 60~kpc. Then, a more robust model of the galaxy can be determined, confirming the decline in density, and a more detailed analysis of the optical depth to the LMC will allow us to convincingly probe the MACHO nature of the dark matter halo. Furthermore, in the near future, with the LSST survey to be performed in the Vera Rubin Observatory, as well as the Nancy Roman Space Telescope, with much better cadence and longer duration, one could extend the range of masses accessible to the microlensing analysis.

\backmatter


\bmhead{Acknowledgments}

JGB thanks G\"unther Hasinger, Alex Drlica-Wagner, Tamara Davis and Fernando Quevedo for thoughtful discussions and useful comments.



\begin{itemize}
\item Funding

JGB acknowledges support from the MICINN [FEDER] research project PID2021-123012NB-C43 and the Spanish Research Agency (Agencia Estatal de Investigaci\'on) through the Grant IFT Centro de Excelencia Severo Ochoa No. CEX2020-001007-S, funded by MCIN/AEI/10.13039/501100011033.

\item Conflict of interest/Competing interests 

The authors declare there are no conflict of interests.

\item Ethics approval 

This work has no ethical conflicts.

\item Consent for publication

Both authors agree on the publication.

\item Availability of data and materials

This work has made use of data from the European Space Agency (ESA) mission
{\it Gaia} (\url{https://www.cosmos.esa.int/gaia}), processed by the {\it Gaia}
Data Processing and Analysis Consortium (DPAC,
\url{https://www.cosmos.esa.int/web/gaia/dpac/consortium}). Funding for the DPAC
has been provided by national institutions, in particular the institutions
participating in the {\it Gaia} Multilateral Agreement.

\item Code availability

We have used a {\tt Mathematica} code with the rotation curve data provided by the authors of Ref.~\citep{Jiao:2023aci}. We can share the code upon request.

\item Authors' contributions

JGB proposed the analysis, created the code, produced the figures and written most of the paper. MRSH has contributed with insightful suggestions on microlensing efficiencies and has written parts of the paper.

\end{itemize}









\bibliography{biblio}


\begin{thebibliography}{29}
\ifx \bisbn   \undefined \def \bisbn  #1{ISBN #1}\fi
\ifx \binits  \undefined \def \binits#1{#1}\fi
\ifx \bauthor  \undefined \def \bauthor#1{#1}\fi
\ifx \batitle  \undefined \def \batitle#1{#1}\fi
\ifx \bjtitle  \undefined \def \bjtitle#1{#1}\fi
\ifx \bvolume  \undefined \def \bvolume#1{\textbf{#1}}\fi
\ifx \byear  \undefined \def \byear#1{#1}\fi
\ifx \bissue  \undefined \def \bissue#1{#1}\fi
\ifx \bfpage  \undefined \def \bfpage#1{#1}\fi
\ifx \blpage  \undefined \def \blpage #1{#1}\fi
\ifx \burl  \undefined \def \burl#1{\textsf{#1}}\fi
\ifx \doiurl  \undefined \def \doiurl#1{\url{https://doi.org/#1}}\fi
\ifx \betal  \undefined \def \betal{\textit{et al.}}\fi
\ifx \binstitute  \undefined \def \binstitute#1{#1}\fi
\ifx \binstitutionaled  \undefined \def \binstitutionaled#1{#1}\fi
\ifx \bctitle  \undefined \def \bctitle#1{#1}\fi
\ifx \beditor  \undefined \def \beditor#1{#1}\fi
\ifx \bpublisher  \undefined \def \bpublisher#1{#1}\fi
\ifx \bbtitle  \undefined \def \bbtitle#1{#1}\fi
\ifx \bedition  \undefined \def \bedition#1{#1}\fi
\ifx \bseriesno  \undefined \def \bseriesno#1{#1}\fi
\ifx \blocation  \undefined \def \blocation#1{#1}\fi
\ifx \bsertitle  \undefined \def \bsertitle#1{#1}\fi
\ifx \bsnm \undefined \def \bsnm#1{#1}\fi
\ifx \bsuffix \undefined \def \bsuffix#1{#1}\fi
\ifx \bparticle \undefined \def \bparticle#1{#1}\fi
\ifx \barticle \undefined \def \barticle#1{#1}\fi
\bibcommenthead
\ifx \bconfdate \undefined \def \bconfdate #1{#1}\fi
\ifx \botherref \undefined \def \botherref #1{#1}\fi
\ifx \url \undefined \def \url#1{\textsf{#1}}\fi
\ifx \bchapter \undefined \def \bchapter#1{#1}\fi
\ifx \bbook \undefined \def \bbook#1{#1}\fi
\ifx \bcomment \undefined \def \bcomment#1{#1}\fi
\ifx \oauthor \undefined \def \oauthor#1{#1}\fi
\ifx \citeauthoryear \undefined \def \citeauthoryear#1{#1}\fi
\ifx \endbibitem  \undefined \def \endbibitem {}\fi
\ifx \bconflocation  \undefined \def \bconflocation#1{#1}\fi
\ifx \arxivurl  \undefined \def \arxivurl#1{\textsf{#1}}\fi
\csname PreBibitemsHook\endcsname

\bibitem[\protect\citeauthoryear{Hawking}{1971}]{Hawking:1971ei}
\begin{barticle}
\bauthor{\bsnm{Hawking}, \binits{S.}}:
\batitle{{Gravitationally collapsed objects of very low mass}}.
\bjtitle{Mon. Not. Roy. Astron. Soc.}
\bvolume{152},
\bfpage{75}
(\byear{1971})
\end{barticle}
\endbibitem

\bibitem[\protect\citeauthoryear{Chapline}{1975}]{Chapline:1975ojl}
\begin{barticle}
\bauthor{\bsnm{Chapline}, \binits{G.F.}}:
\batitle{{Cosmological effects of primordial black holes}}.
\bjtitle{Nature}
\bvolume{253}(\bissue{5489}),
\bfpage{251}
(\byear{1975})
\end{barticle}
\endbibitem

\bibitem[\protect\citeauthoryear{Carr et~al.}{2024}]{Carr:2023tpt}
\begin{barticle}
\bauthor{\bsnm{Carr}, \binits{B.}},
\bauthor{\bsnm{Clesse}, \binits{S.}},
\bauthor{\bsnm{Garcia-Bellido}, \binits{J.}},
\bauthor{\bsnm{Hawkins}, \binits{M.}},
\bauthor{\bsnm{Kuhnel}, \binits{F.}}:
\batitle{{Observational evidence for primordial black holes: A positivist
  perspective}}.
\bjtitle{Phys. Rept.}
\bvolume{1054},
\bfpage{1}--\blpage{68}
(\byear{2024})
{\href{https://arxiv.org/abs/2306.03903}{{arXiv:2306.03903}}}
{[astro-ph.CO]}
\end{barticle}
\endbibitem

\bibitem[\protect\citeauthoryear{Bagui
  et~al.}{2023}]{LISACosmologyWorkingGroup:2023njw}
\begin{botherref}
\oauthor{\bsnm{Bagui}, \binits{E.}}, et al.:
{Primordial black holes and their gravitational-wave signatures}
(2023)
{\href{https://arxiv.org/abs/2310.19857}{{arXiv:2310.19857}}}
{[astro-ph.CO]}
\end{botherref}
\endbibitem

\bibitem[\protect\citeauthoryear{Garc\'ia-Bellido
  et~al.}{1996}]{Garcia-Bellido:1996mdl}
\begin{barticle}
\bauthor{\bsnm{Garc\'ia-Bellido}, \binits{J.}},
\bauthor{\bsnm{Linde}, \binits{A.D.}},
\bauthor{\bsnm{Wands}, \binits{D.}}:
\batitle{{Density perturbations and black hole formation in hybrid inflation}}.
\bjtitle{Phys. Rev. D}
\bvolume{54},
\bfpage{6040}
(\byear{1996})
{\href{https://arxiv.org/abs/astro-ph/9605094}{{arXiv:astro-ph/9605094}}}
\end{barticle}
\endbibitem

\bibitem[\protect\citeauthoryear{Jedamzik}{1997}]{Jedamzik:1996mr}
\begin{barticle}
\bauthor{\bsnm{Jedamzik}, \binits{K.}}:
\batitle{{Primordial black hole formation during the QCD epoch}}.
\bjtitle{Phys. Rev. D}
\bvolume{55},
\bfpage{5871}
(\byear{1997})
{\href{https://arxiv.org/abs/astro-ph/9605152}{{arXiv:astro-ph/9605152}}}
\end{barticle}
\endbibitem

\bibitem[\protect\citeauthoryear{Paczynski}{1986}]{Paczynski:1985jf}
\begin{barticle}
\bauthor{\bsnm{Paczynski}, \binits{B.}}:
\batitle{{Gravitational microlensing by the galactic halo}}.
\bjtitle{Astrophys. J.}
\bvolume{304},
\bfpage{1}
(\byear{1986})
\end{barticle}
\endbibitem

\bibitem[\protect\citeauthoryear{Alcock et~al.}{1996}]{MACHO:1995udp}
\begin{barticle}
\bauthor{\bsnm{Alcock}, \binits{C.}}, \betal:
\batitle{{The MACHO project first year LMC results: The Microlensing rate and
  the nature of the galactic dark halo}}.
\bjtitle{Astrophys. J.}
\bvolume{461},
\bfpage{84}
(\byear{1996})
{\href{https://arxiv.org/abs/astro-ph/9506113}{{arXiv:astro-ph/9506113}}}
\end{barticle}
\endbibitem

\bibitem[\protect\citeauthoryear{Alcock et~al.}{2000}]{MACHO:2000qbb}
\begin{barticle}
\bauthor{\bsnm{Alcock}, \binits{C.}}, \betal:
\batitle{{The MACHO project: Microlensing results from 5.7 years of LMC
  observations}}.
\bjtitle{Astrophys. J.}
\bvolume{542},
\bfpage{281}
(\byear{2000})
{\href{https://arxiv.org/abs/astro-ph/0001272}{{arXiv:astro-ph/0001272}}}
\end{barticle}
\endbibitem

\bibitem[\protect\citeauthoryear{Hawkins}{2015}]{Hawkins:2015uja}
\begin{barticle}
\bauthor{\bsnm{Hawkins}, \binits{M.R.S.}}:
\batitle{{A new look at microlensing limits on dark matter in the Galactic
  halo}}.
\bjtitle{Astron. Astrophys.}
\bvolume{575},
\bfpage{107}
(\byear{2015})
{\href{https://arxiv.org/abs/1503.01935}{{arXiv:1503.01935}}}
\end{barticle}
\endbibitem

\bibitem[\protect\citeauthoryear{Tisserand et~al.}{2007}]{EROS-2:2006ryy}
\begin{barticle}
\bauthor{\bsnm{Tisserand}, \binits{P.}}, \betal:
\batitle{{Limits on the Macho Content of the Galactic Halo from the EROS-2
  Survey of the Magellanic Clouds}}.
\bjtitle{Astron. Astrophys.}
\bvolume{469},
\bfpage{387}
(\byear{2007})
{\href{https://arxiv.org/abs/astro-ph/0607207}{{arXiv:astro-ph/0607207}}}
\end{barticle}
\endbibitem

\bibitem[\protect\citeauthoryear{Wyrzykowski et~al.}{2011}]{Wyrzykowski:2011tr}
\begin{barticle}
\bauthor{\bsnm{Wyrzykowski}, \binits{L.}}, \betal:
\batitle{{The OGLE View of Microlensing towards the Magellanic Clouds. IV.
  OGLE-III SMC Data and Final Conclusions on MACHOs}}.
\bjtitle{Mon. Not. Roy. Astron. Soc.}
\bvolume{416},
\bfpage{2949}
(\byear{2011})
{\href{https://arxiv.org/abs/1106.2925}{{arXiv:1106.2925}}}
\end{barticle}
\endbibitem

\bibitem[\protect\citeauthoryear{Clemens}{1985}]{Clemens:1985}
\begin{barticle}
\bauthor{\bsnm{Clemens}, \binits{D.P.}}:
\batitle{{Massachusetts-Stony Brook Galactic plane CO survey: the galactic disk
  rotation curve}}.
\bjtitle{Astrophys. J.}
\bvolume{295},
\bfpage{422}
(\byear{1985})
\end{barticle}
\endbibitem

\bibitem[\protect\citeauthoryear{Xue et~al.}{2008}]{Xue:2008nmx}
\begin{barticle}
\bauthor{\bsnm{Xue}, \binits{X.X.}}, \betal:
\batitle{{The Milky Way's Circular Velocity Curve to 60 kpc and an Estimate of
  the Dark Matter Halo Mass from Kinematics of \textasciitilde{}2400 SDSS Blue
  Horizontal Branch Stars}}.
\bjtitle{Astrophys. J.}
\bvolume{684},
\bfpage{1143}
(\byear{2008})
{\href{https://arxiv.org/abs/0801.1232}{{arXiv:0801.1232}}}
\end{barticle}
\endbibitem

\bibitem[\protect\citeauthoryear{Sofue}{2013}]{Sofue:2013kja}
\begin{barticle}
\bauthor{\bsnm{Sofue}, \binits{Y.}}:
\batitle{{Rotation Curve and Mass Distribution in the Galactic Center --- From
  Black Hole to Entire Galaxy ---}}.
\bjtitle{Publ. Astron. Soc. Jap.}
\bvolume{65},
\bfpage{118}
(\byear{2013})
{\href{https://arxiv.org/abs/1307.8241}{{arXiv:1307.8241}}}
\end{barticle}
\endbibitem

\bibitem[\protect\citeauthoryear{Bhattacharjee
  et~al.}{2014}]{Bhattacharjee:2013exa}
\begin{barticle}
\bauthor{\bsnm{Bhattacharjee}, \binits{P.}},
\bauthor{\bsnm{Chaudhury}, \binits{S.}},
\bauthor{\bsnm{Kundu}, \binits{S.}}:
\batitle{{Rotation Curve of the Milky Way out to $\sim$ 200 kpc}}.
\bjtitle{Astrophys. J.}
\bvolume{785},
\bfpage{63}
(\byear{2014})
{\href{https://arxiv.org/abs/1310.2659}{{arXiv:1310.2659}}}
\end{barticle}
\endbibitem

\bibitem[\protect\citeauthoryear{Calcino et~al.}{2018}]{Calcino:2018mwh}
\begin{barticle}
\bauthor{\bsnm{Calcino}, \binits{J.}},
\bauthor{\bsnm{Garc\'ia-Bellido}, \binits{J.}},
\bauthor{\bsnm{Davis}, \binits{T.M.}}:
\batitle{{Updating the MACHO fraction of the Milky Way dark halo with improved
  mass models}}.
\bjtitle{Mon. Not. Roy. Astron. Soc.}
\bvolume{479}(\bissue{3}),
\bfpage{2889}
(\byear{2018})
{\href{https://arxiv.org/abs/1803.09205}{{arXiv:1803.09205}}}
\end{barticle}
\endbibitem

\bibitem[\protect\citeauthoryear{Huang et~al.}{2016}]{Huang:2016}
\begin{barticle}
\bauthor{\bsnm{Huang}, \binits{Y.}}, \betal:
\batitle{{The Milky Way's rotation curve out to 100 kpc and its constraint on
  the Galactic mass distribution}}.
\bjtitle{Mon. Not. Roy. Astron. Soc.}
\bvolume{463},
\bfpage{2623}
(\byear{2016})
{\href{https://arxiv.org/abs/1604.01216}{{arXiv:1604.01216}}}
\end{barticle}
\endbibitem

\bibitem[\protect\citeauthoryear{Ou et~al.}{2023}]{Ou:2023adg}
\begin{botherref}
\oauthor{\bsnm{Ou}, \binits{X.}},
\oauthor{\bsnm{Eilers}, \binits{A.-C.}},
\oauthor{\bsnm{Necib}, \binits{L.}},
\oauthor{\bsnm{Frebel}, \binits{A.}}:
{The dark matter profile of the Milky Way inferred from its circular velocity
  curve}
(2023)
{\href{https://arxiv.org/abs/2303.12838}{{arXiv:2303.12838}}}
\end{botherref}
\endbibitem

\bibitem[\protect\citeauthoryear{Jiao et~al.}{2023}]{Jiao:2023aci}
\begin{barticle}
\bauthor{\bsnm{Jiao}, \binits{Y.}},
\bauthor{\bsnm{Hammer}, \binits{F.}},
\bauthor{\bsnm{Wang}, \binits{H.}},
\bauthor{\bsnm{Wang}, \binits{J.}},
\bauthor{\bsnm{Amram}, \binits{P.}},
\bauthor{\bsnm{Chemin}, \binits{L.}},
\bauthor{\bsnm{Yang}, \binits{Y.}}:
\batitle{{Detection of the Keplerian decline in the Milky Way rotation curve}}.
\bjtitle{Astron. Astrophys.}
\bvolume{678},
\bfpage{208}
(\byear{2023})
{\href{https://arxiv.org/abs/2309.00048}{{arXiv:2309.00048}}}
\end{barticle}
\endbibitem

\bibitem[\protect\citeauthoryear{Wang et~al.}{2023}]{Wang:2023}
\begin{barticle}
\bauthor{\bsnm{Wang}, \binits{H.-F.}},
\bauthor{\bsnm{Chrob\'akov\'a}, \binits{Z.}},
\bauthor{\bsnm{L\'opez-Corredoira}, \binits{M.}},
\bauthor{\bsnm{Sylos~Labini}, \binits{F.}}:
\batitle{{Mapping the Milky Way Disk with Gaia DR3: 3D Extended Kinematic Maps
  and Rotation Curve to 30 kpc}}.
\bjtitle{Astrophys. J.}
\bvolume{942},
\bfpage{12}
(\byear{2023})
\end{barticle}
\endbibitem

\bibitem[\protect\citeauthoryear{Sylos~Labini et~al.}{2023}]{SylosLabini:2023}
\begin{barticle}
\bauthor{\bsnm{Sylos~Labini}, \binits{F.}},
\bauthor{\bsnm{Chrob\'akov\'a}, \binits{Z.}},
\bauthor{\bsnm{Capuzzo-Dolcetta}, \binits{R.}},
\bauthor{\bsnm{L\'opez-Corredoira}, \binits{M.}}:
\batitle{{Mass Models of the Milky Way and Estimation of Its Mass from the Gaia
  DR3 Data Set}}.
\bjtitle{Astrophys. J.}
\bvolume{945},
\bfpage{3}
(\byear{2023})
\end{barticle}
\endbibitem

\bibitem[\protect\citeauthoryear{Vallenari et~al.}{2023}]{DR3}
\begin{barticle}
\bauthor{\bsnm{Vallenari}, \binits{A.}}, \betal:
\batitle{{Gaia Data Release 3. Summary of the content and survey properties}}.
\bjtitle{Astron. Astrophys.}
\bvolume{674},
\bfpage{1}
(\byear{2023})
{\href{https://arxiv.org/abs/2208.00211}{{arXiv:2208.00211}}}
\end{barticle}
\endbibitem

\bibitem[\protect\citeauthoryear{Bennett}{2005}]{Bennett:2005at}
\begin{barticle}
\bauthor{\bsnm{Bennett}, \binits{D.P.}}:
\batitle{{Large Magellanic Cloud microlensing optical depth with imperfect
  event selection}}.
\bjtitle{Astrophys. J.}
\bvolume{633},
\bfpage{906}
(\byear{2005})
{\href{https://arxiv.org/abs/astro-ph/0502354}{{arXiv:astro-ph/0502354}}}
\end{barticle}
\endbibitem

\bibitem[\protect\citeauthoryear{Alcock et~al.}{2001}]{Alcock:2000zv}
\begin{barticle}
\bauthor{\bsnm{Alcock}, \binits{C.}}, \betal:
\batitle{{The MACHO project Hubble Space Telescope follow-up: preliminary
  results on the location of the large magellanic cloud microlensing source
  stars}}.
\bjtitle{Astrophys. J.}
\bvolume{552},
\bfpage{582}
(\byear{2001})
{\href{https://arxiv.org/abs/astro-ph/0008282}{{arXiv:astro-ph/0008282}}}
\end{barticle}
\endbibitem

\bibitem[\protect\citeauthoryear{Byrnes et~al.}{2018}]{Byrnes:2018clq}
\begin{barticle}
\bauthor{\bsnm{Byrnes}, \binits{C.T.}},
\bauthor{\bsnm{Hindmarsh}, \binits{M.}},
\bauthor{\bsnm{Young}, \binits{S.}},
\bauthor{\bsnm{Hawkins}, \binits{M.R.S.}}:
\batitle{{Primordial black holes with an accurate QCD equation of state}}.
\bjtitle{JCAP}
\bvolume{08},
\bfpage{041}
(\byear{2018})
{\href{https://arxiv.org/abs/1801.06138}{{arXiv:1801.06138}}}
\end{barticle}
\endbibitem

\bibitem[\protect\citeauthoryear{Refsdal and Stabell}{1991}]{Refsdal:1991}
\begin{barticle}
\bauthor{\bsnm{Refsdal}, \binits{S.}},
\bauthor{\bsnm{Stabell}}:
\batitle{{Gravitational micro-lensing for large sources}}.
\bjtitle{Astron. Astrophys.}
\bvolume{250},
\bfpage{62}
(\byear{1991})
\end{barticle}
\endbibitem

\bibitem[\protect\citeauthoryear{Alcock et~al.}{2001}]{MACHO:2000bzs}
\begin{barticle}
\bauthor{\bsnm{Alcock}, \binits{C.}}, \betal:
\batitle{{The MACHO project: microlensing detection efficiency}}.
\bjtitle{Astrophys. J. Suppl.}
\bvolume{136},
\bfpage{439}
(\byear{2001})
{\href{https://arxiv.org/abs/astro-ph/0003392}{{arXiv:astro-ph/0003392}}}
\end{barticle}
\endbibitem

\bibitem[\protect\citeauthoryear{Carr et~al.}{2021}]{Carr:2019kxo}
\begin{barticle}
\bauthor{\bsnm{Carr}, \binits{B.}},
\bauthor{\bsnm{Clesse}, \binits{S.}},
\bauthor{\bsnm{Garc\'\i{}a-Bellido}, \binits{J.}},
\bauthor{\bsnm{K\"uhnel}, \binits{F.}}:
\batitle{{Cosmic conundra explained by thermal history and primordial black
  holes}}.
\bjtitle{Phys. Dark Univ.}
\bvolume{31},
\bfpage{100755}
(\byear{2021})
{\href{https://arxiv.org/abs/1906.08217}{{arXiv:1906.08217}}}
\end{barticle}
\endbibitem

\end{thebibliography}

\end{document}